\documentclass[journal=jpcc,manuscript=article]{achemso}
\usepackage{graphicx}
\usepackage{dcolumn}
\usepackage{bm}
\title{Thermal Expansion of Ni-Ti-Sn Heusler and half-Heusler Materials From First Principles Calculations and Experiments}
\author{P. Hermet, R.M. Ayral, E. Theron, P. G. Yot, F. Salles, M. Tillard and P. Jund}
\affiliation{Institut Charles Gerhardt Montpellier, UMR 5253 CNRS-UM2-ENSCM-UM1, Universit\'e Montpellier 2, Place E. Bataillon, 34095 Montpellier C\'edex 5, France}
\email{pjund@um2.fr}
\begin{document}
\begin{abstract}
We coupled first principles calculations and the quasiharmonic
approximation combined with experiments (X-Ray diffraction and
dilatometry measurements) to determine the thermal properties of
NiTiSn (half-Heusler) and Ni$_2$TiSn (Heusler) compounds. These
properties are important especially if they are to be used in
thermoelectric applications. First, the calculation of their mode
Gruneisen parameter shows that it is positive throughout 
the first Brillouin zone. This suggests that these
compounds undergo a regular thermal expansion. Then, the calculation
of the Ni$_2$TiSn thermal expansion shows an excellent agreement, even
in the high temperature range, with our high energy powder X-Ray diffraction
measurements (ESRF) and dilatometry experiments. In the case of
NiTiSn, this agreement is less impressive. This is due to stronger
phonon-phonon interactions that are not considered within the
quasiharmonic approximation, but also to the difficulty of making
high-quality NiTiSn samples. Finally, the constant-pressure and
constant-volume heat capacities have been calculated for both
compounds and compared with the experimental data reported in the
literature. In particular, we have decomposed the constant-volume heat capacity of Ni$_2$TiSn into a purely electronic and a phonon-mediated contribution, and we discuss each of them.   
\end{abstract} 


\section{I. INTRODUCTION}
Heusler and half-Heusler compounds are the archetypes for multifunctional materials and their versatility increases constantly \cite{properties}. They generally crystallize into cubic structures (Fm$\bar3$m for the Heusler and F$\bar 4$3m for the half-Heusler are the usual space groups \cite{pearson}), which increases their theoretical attractivity. In addition, they are made of environmental friendly elements which permits to use them in practical applications. In this paper, our work is focused on the Ni$_2$TiSn Heusler and NiTiSn half-Heusler compounds which have good thermoelectric properties, properties that are of interest to us. In previous studies, we have tackled through {\em ab initio} simulations the electronic properties \cite{colinet} and the thermodynamic/mechanical \cite{patNiTiSn} properties of these compounds since a comprehensive study of these materials is still missing even though it is necessary if one wants to use them in real thermoelectric devices. \\
The thermoelectric effect is a direct conversion of a temperature gradient into an electric current {\it via} the Seebeck effect. Thus, thermoelectric generators contribute to sustainability through scavenging of waste heat or heat sources \cite{snyder}. This implies that an accurate knowledge of the behavior of the thermoelectric materials as a function of temperature is necessary. Such a knowledge is missing for the Heusler type materials considered in our study. For the half-Heusler,  Wee {\it et al.}~\cite{Wee} have calculated the Gr\"uneisen parameters of the acoustic phonon modes to do a rough estimate of its lattice thermal conductivity, and Jung {\it et al.}~\cite{Jung} have measured its thermal expansion up to its decomposition point. 

We propose here to couple the quasiharmonic approximation and the
density functional perturbation theory to study the thermodynamic
properties of NiTiSn and Ni$_2$TiSn for temperatures up to 1500~K
(close to the melting temperature of NiTiSn). For this purpose,  we
first compute the mode Gruneisen parameter of the acoustic and optical
phonon modes and the constant-volume heat capacity for both
compounds. These quantities are indeed of special interest to clearly
understand their thermal expansion mechanism. Then, we derive their
constant-pressure heat capacity. In the case of Ni$_2$TiSn, the latter
is decomposed into a purely electronic and a phonon-mediated
contribution, and we discuss each of these contributions. 
Finally, our calculations will be compared with experimental data. In
this context, we measured the thermal expansion of Ni$_2$TiSn for
which no such data exist in the literature using temperature dependent
powder X-Ray diffraction measurements from 80 to 673~K on laboratory equipment and
at the European Synchrotron Research Facility (ESRF) as well as
dilatometry experiments up to 873~K. Although the experimental thermal
expansion of NiTiSn has been reported by Jung {\it et
  al.}~\cite{Jung}, we revisited this work using new X-ray
measurements for the sake of completeness and to compare our data with
the ones already published. 

The paper is organized as follows: in section two, we present the experimental conditions whereas we detail the theoretical framework in section three. In section four, we show and discuss the results, while section five draws the major conclusions of our work.

\section{II. EXPERIMENTAL CONDITIONS}
The elements (Ti (99,7\%, 90$\mu$m), Ni (99,9\% , 5 $\mu$m) and Sn (99,85\% , 90 $\mu$m) were weighed
in quantities corresponding to the desired stoichiometry. For the
synthesis of Ni$_2$TiSn, the elements were mixed and compacted at 375 MPa during 30 minutes under primary vacuum. Ingots 
of Ni$_2$TiSn were then prepared by arc melting. 
The NiTiSn ingots prepared by arc melting as well were sealed into silica tubes and subsequently annealed at 1073K for ten days in order to improve their quality.
The resultant materials were investigated by X-ray diffraction (XRD)
using the Cu K$_\alpha$-radiation in the 2$\theta$ range from 20 to 60$^\circ$. The observation of the morphology of the materials coupled to the quantitative chemical analysis was performed on a scanning electron microscope FEI Quanta 200, resolution 3 nm, vacuum, 30kV for the secondary electrons and 4 nm for the backscattered electrons, coupled with a microprobe EDS (Energy Dispersive X-ray Spectroscopy) Oxford Instrument XMax with a detector of 50 mm$^2$. In each case, the desired phase was obtained. 

Concerning the Ni$_2$TiSn material, measurements were made both at low
and high temperatures. For low temperature experiments (performed at the European Synchrotron Radiation Facility - ESRF,
France), a fine powder of Ni$_2$TiSn was loaded into a glass capillary of 0.5~mm diameter. The
capillary was sealed and cooled from room temperature to 80 K and then heated 
at a rate of 30~K/hour up to 250 K, while synchrotron
powder-diffraction data were collected in situ. 
X-Ray diffraction was performed on the Swiss-Norwegian Beam Line
(BM01A) at the ESRF. All the diffraction patterns were collected using a
monochromatic beam with the wavelength of 0.70814~\AA\ and a PILATUS 2M
detector. The sample-detector distance (343.71 mm) and parameters of
the detector were calibrated using NIST standard
LaB$_6$. Two-dimensional diffraction images were integrated using the
Fit2D software \cite{ham}. The temperature was controlled with an Oxford Cryostream 700+. 
During each collection time (120 s per image) the capillary was rotated by 60$^\circ$ in the same angular interval. 
X-ray measurements as a function of temperature in the temperature
range 293 to 673 K were also realized for Ni$_2$TiSn (performed at the ``Laboratoire de Physique du Solide'' (LPS) in Paris-Orsay) with a heating rate of 5~K/min. The X-ray set-up consists of a
diffractometer in the normal-beam geometry. The sealed silica
capillary containing the powder is fixed on a holder rotating
perpendicularly to the incident beam. It is mounted on a rotating
anode generator (copper anode: $\lambda$ = 1.542~\AA) equipped with a
doubly curved graphite monochromator. The sample-holder was enclosed
into a furnace. The diffraction pattern was recorded on an image plate
detector (MAR 345) located at a distance of 425~mm from the sample and
tilted of 20$^\circ$ from the incident beam direction.
Finally we measured also the thermal expansion at the PRIME Verre company (Montpellier) 
with a ADAMEL-LHOMARGY model DI.10.2 dilatometer for rectangular-shaped samples cut from the
ingots between room temperature and 873~K under air, with a heating
rate of 3 K/min. From the obtained thermal expansion curves, the average volumetric
thermal expansion coefficient was calculated with uncertainties of 0.1~MK$^{-1}$.

For the NiTiSn material, X-Ray analyses as a function of temperature
were performed at the ICGM (Montpellier) using a temperature PANalytical X'Pert Pro Philips
apparatus. The unit was equipped with an Anton Paar HTK 1200 high
temperature chamber to work under vacuum (pressure 7.10$^{-4}$ mbar)
and for temperatures between 293 and 673~K. The NiTiSn powders were
first mixed with $\alpha$ alumina in an ethanol solution in order to
obtain an homogeneous repartition of the powder in the sample holder.
$\alpha$ alumina was used in these experiments as reference because the dilatation
coefficient of this material as a function of temperature is
well-known and served thus to correct the dilatation phenomena
inherent to the apparatus.
Acquisition was realized between 20 and 90$^\circ$ (2$\theta$) with a
step size 0.0130$^\circ$ and a scanning time of 600~s. Dwell time was
kept at 10~min for stabilizing 
the temperature before recording the X-rays. 

\section{III. THEORETICAL FRAMEWORK}
  \subsection{A. Thermal expansion coefficient}

The equilibrium volume of a crystal at a given temperature $T$ and in the absence of any applied pressure is obtained by minimizing the Helmholtz free energy  with respect to all possible internal degrees of freedom:
\begin{equation}
\label{FreeE}
F(V, T) = E_0(V) + F_{el}(V, T) + F_{ph} (V, T), 
\end{equation}
where $V$ is the unit cell volume, $E_0$ is the ground state ($T=$0~K) total energy of the crystal, $F_{el}$ is the electronic free energy contribution, and $F_{ph}$ is the vibrational free energy which comes from the phonon contribution. In the quasiharmonic approximation (QHA), the latter term is written as follows~\cite{Maradudin,PatAgCo}:
\begin{equation}
\label{Fph}
F_{ph}(V, T) = \frac{\hbar}{2}\sum_{j,\mathbf{q}}  \omega(j,\mathbf{q},V)  + k_BT \sum_{j,\mathbf{q}} ln \left[ 1- exp\left( - \frac{\hbar\omega(j,\mathbf{q},V)}{k_BT} \right) \right],
\end{equation}
where the sums run over all allowed wavevectors $\mathbf{q}$ in the first Brillouin zone and over all phonon branches $j$, $k_B$ is the Boltzmann constant, $\hbar$ is the reduced Planck constant, and $\omega(j,\mathbf{q},V)$ is the frequency of the phonon with wavevector $\mathbf{q} $ in branch $j$, evaluated at constant-volume $V$. The first and second terms of \ref{Fph} represent the zero-point and thermal energies of the phonons, respectively. $F_{ph}$ can be obtained by calculating the phonon dispersion relations and the corresponding phonon density of states (DOS) from density functional perturbation theory. The QHA assumes that the phonon frequency is only a function of volume and is temperature-independent. Thus, the electron-phonon coupling is neglected.

For the electronic contribution to the free energy, $F_{el}=E_{el}-TS_{el}$, the electronic energy due to thermal electronic excitations is given by~\cite{Wasserman}:
\begin{equation}
E_{el}(V, T) = \int_0^\infty n(\varepsilon,V) f(\varepsilon)\varepsilon d\varepsilon - \int_0^{\varepsilon_F} n(\varepsilon,V) \varepsilon d\varepsilon,
\end{equation}
where  $n(\varepsilon,V)$, $f(\varepsilon)$, and $\varepsilon_F$ represent the electronic density of states, the Fermi-Dirac distribution, and the Fermi energy, respectively. The electronic entropy is formulated as:
\begin{equation}
S_{el}(V, T) =-k_B\int_0^\infty n(\varepsilon,V)  [f(\varepsilon) ln f(\varepsilon)+(1-f(\varepsilon))ln(1-f(\varepsilon))]d\varepsilon.
\end{equation}
These contributions are determined from the calculated electronic DOS.

By definition, the volumetric~\cite{Note} thermal expansion coefficient of a crystal is given by~\cite{Barron}:
\begin{equation}
\alpha_V=\left(\frac{\partial ln V}{\partial T}\right)_P,
\end{equation}
where the subscript $P$ implies that the temperature derivative is taken at constant pressure. For cubic symmetry, the minimization of \ref{FreeE} with respect to the volume leads to an alternative expression for $\alpha_V$  composed of the sum of a vibrational term ($\alpha^{ph}$) and an electronic term ($\alpha^{el}$):
\begin{eqnarray} 
\alpha_V & = &\alpha^{ph}+\alpha^{el}  \\
 & = &\frac{1}{BV} \sum_{j,\mathbf{q}} \gamma_j(\mathbf{q}) C_v^{ph}(j,\mathbf{q}) + \frac{2}{3BV}C_v^{el},
\end{eqnarray} 
where $B$ is the bulk modulus and $\gamma_j(\mathbf{q})$ is the mode Gr\"uneisen parameter defined as:
\begin{equation}
\label{grun}
\gamma_j(\mathbf{q})=-\left(\frac{\partial ln[\omega(j,\mathbf{q},V)]}{\partial ln V} \right)_0.
\end{equation}
Here, the subscript "0" indicates a quantity taken at the ground state lattice parameter. The vibrational specific heat at constant-volume is obtained using the calculated phonon DOS as follows:
\begin{equation}
\label{cvph}
C^{ph}_v = k_B \sum_{j,\mathbf{q}} \left( \frac{\hbar\omega(j,\mathbf{q})}{2k_BT} \right)^2 csch^2 \left( \frac{\hbar\omega(j,\mathbf{q})}{2k_BT}\right),
\end{equation}  
and the electronic specific heat at constant-volume can be obtained from~\cite{Wasserman}:
\begin{equation}
\label{cvel}
C^{el}_v = T\left( \frac{\partial S_{el}}{\partial T}\right)_V.
\end{equation}  
The specific heat at constant-pressure can be determined by using the relation:
\begin{equation}
\label{cpcv}
C_p-C_v = \alpha_V^2(T).B(T).V(T).T,
\end{equation}  
where $C_v=C_v^{ph}+C_v^{el}$ is the total specific heat at constant-volume.

\subsection{B. Computational details}

Density functional theory (DFT) based calculations were performed using the ABINIT package~\cite{ABINIT} and the generalized gradient approximation (GGA) parametrized by Perdew, Burke and Ernzerhof (PBE)~\cite{PBE}. The all-electron potentials were replaced by norm-conserving pseudopotentials generated according to the Troullier and Martins scheme~\cite{TM}. Ni($3d^8$, $4s^2$), Ti($3d^2$, $4s^2$) and Sn($5s^2$, $5p^2$)-electrons were considered as valence states in the construction of the pseudopotentials. The electronic wavefunctions were expanded in plane-waves up to a kinetic energy cutoff of 65~Ha. Integrals over the Brillouin zone were approximated by sums over a 8$\times$8$\times$8 mesh of special $k$-points according to the Monkhorst and Pack scheme~\cite{Monkhorst}. A Fermi-Dirac scheme with a smearing width equal to 0.01~Ha was used for the metallic occupation of Ni$_2$TiSn. Phonon dispersion curves were interpolated over a  4 $\times$ 4 $\times$ 4 $q$-points grid according to the scheme described by Gonze {\it et al.} \cite{Gonze}, whereas a denser 120 $\times$ 120  $\times$ 120 grid was employed for the calculation of the  thermodynamic properties.

\section{IV. RESULTS AND DISCUSSION}

\subsection{A. Structural properties}
\begin{figure}
\begin{center}
\includegraphics[width=15cm]{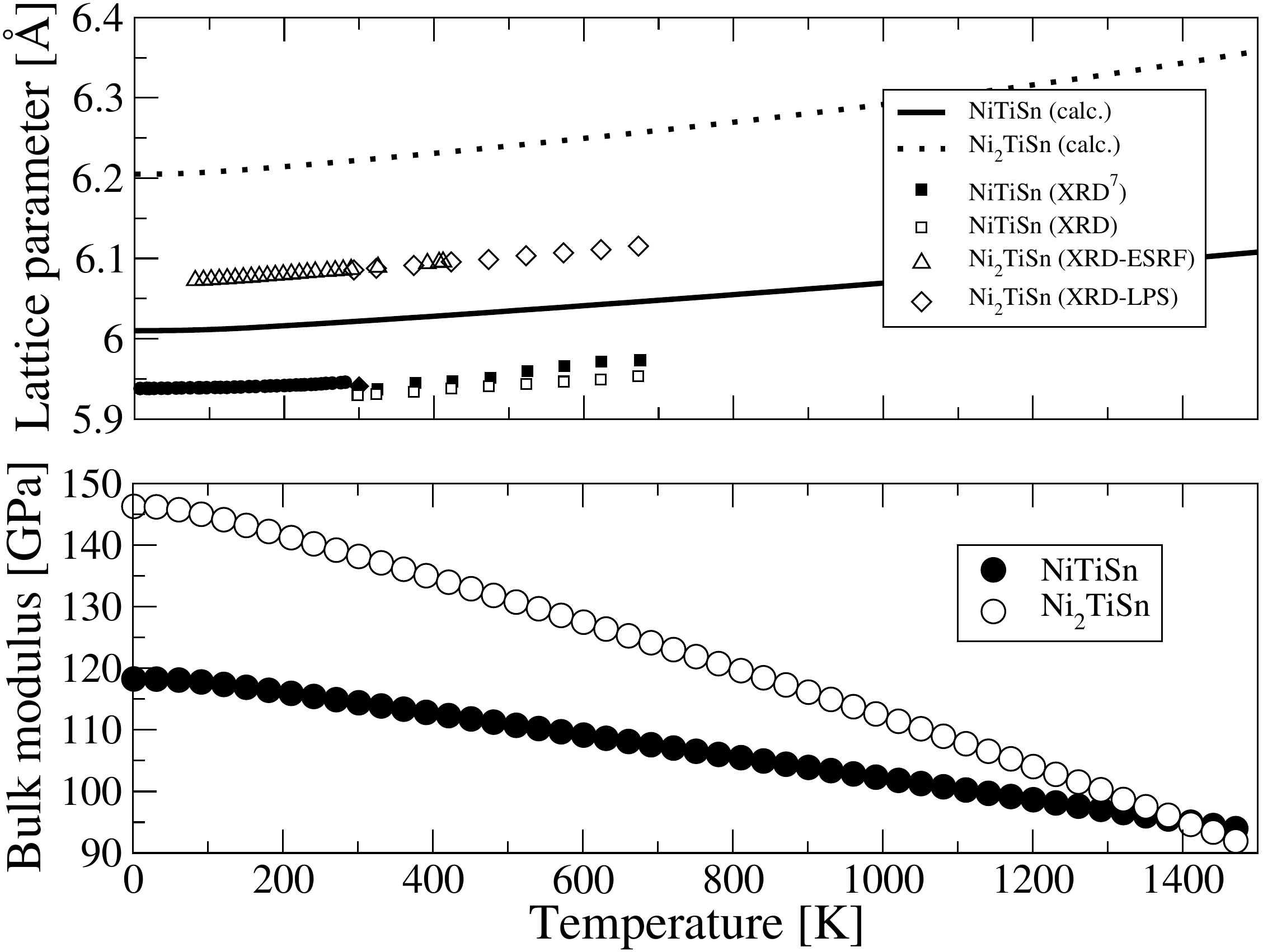}
\caption{Temperature dependence of the lattice parameter (top) and the bulk modulus (bottom) for Ni$_2$TiSn and NiTiSn. Full squares represent the experimental data from Jung {\it et al.}~\cite{Jung}.}
\label{BM}
\end{center}
\end{figure}

In order to get the temperature dependence of the lattice parameters of NiTiSn and Ni$_2$TiSn, we have calculated the total free energy at temperature points from 0 to 1500~K with a step of 15~K and for five volumes. At each temperature point, the equilibrium volume and the isothermal bulk modulus, $B(T)=V \left(\frac{\partial^2 F}{\partial V^2}\right)_T$, is obtained minimizing the free energy from a Birch-Murnaghan equation of state~\cite{BM1,BM2}. Results are shown in \ref{BM} with the experimental ones. Both for the calculation and the experiment, the equilibrium lattice parameters (top panel) show a regular thermal expansion: they increase with increasing temperature. The lattice parameters are sligthly overestimated by our calculations within 1--2\% with respect to the experiments, as usual with GGA exchange--correlation functionals~\cite{Koch}. The dependence of the isothermal bulk modulus as a function of temperature is displayed in the bottom panel. For both compounds, it is almost a linearly decreasing function as the temperature increases. However, the slope of these lines is larger for the metallic compound than for the semiconducting one. As a consequence, these lines intersect near the melting temperature of NiTiSn ($T_m = $1453~K~\cite{Jung}). Below this temperature, the NiTiSn bulk modulus remains the lowest. This is due to the Ni-vacancies which make its structure softer to an hydrostatic pressure.  To our knowledge, there is no experimental data to check our predictions.

\subsection{B. Mode Gr\"uneisen parameters} 
\begin{figure}
\begin{center}
\includegraphics[width=15cm]{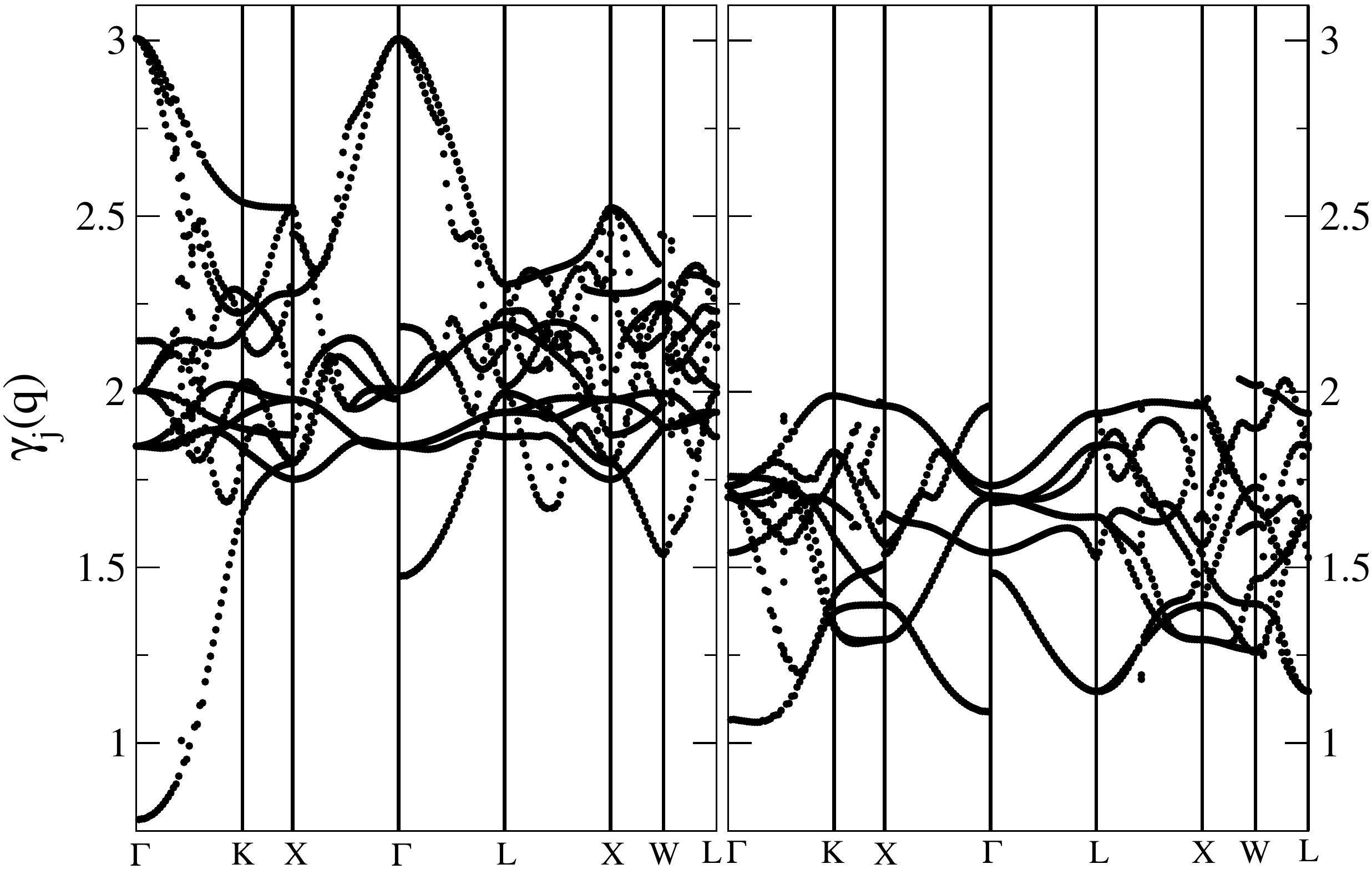}
\caption{Calculated dispersion curves of the mode Gr\"uneisen parameter ($\gamma_j(\mathbf{q})$) for Ni$_2$TiSn (left) and NiTiSn (right) along some high symmetry lines.}
\label{Gru}
\end{center}
\end{figure}

\ref{Gru} reports the mode Gr\"uneisen parameter of NiTiSn and
Ni$_2$TiSn as defined by \ref{grun} along some high symmetry
directions. This calculation requires the knowledge of the phonon
dispersion curves at two additional unit cell volumes. They are
derived from the equilibrium volume by straining it by $\pm$ 3\%, and
reoptimizing the atomic positions. The dispersion curves are
discontinuous at the zone-center as a consequence of the polarization
dependence of the sound velocities. Indeed, because of the vanishing
of the acoustic frequencies at the $\Gamma$-point, the dispersions of
the acoustic branches are discontinuous at the Brillouin zone center,
and the value of the Gr\"uneisen parameters in the limit q
$\rightarrow 0$ depends, for such modes, on the direction of $\mathbf{q}$. We
observe that the Gr\"uneisen parameters are positive throughout the
Brillouin zone for all branches. This suggests from Eq. (7) that there
is no anomalous negative thermal expansion at low temperatures in these compounds since
the heat capacity and the bulk modulus are always positive. While no clear dominant mode Gr\"uneisen parameter is observed for NiTiSn, some branches have a high Gr\"uneisen parameter at the zone-center for Ni$_2$TiSn. This contributes to the increase of the overall Gr\"uneisen parameter of Ni$_2$TiSn at low temperature, since at these temperatures only the very low-frequency phonons contribute to the summation in \ref{grun}, whereas high-frequency phonons decay exponentially. The overall Gr\"uneisen parameter of Ni$_2$TiSn is $\bar\gamma=$2.08, higher than the one of NiTiSn ($\bar\gamma=$1.60). The most positive Gr\"uneisen parameters in Ni$_2$TiSn are located near the $\Gamma$-point and their frequencies are below 150~cm$^{-1}$. These modes mainly involve motions of Ni-atoms (see for instance the case of the T$_{2g}$ Raman mode calculated at 122~cm$^{-1}$ in Ref.~\cite{patNiTiSn}). 

\subsection{C. Thermal expansion coefficients} 
\begin{figure}
\begin{center}
 \includegraphics[width=15cm]{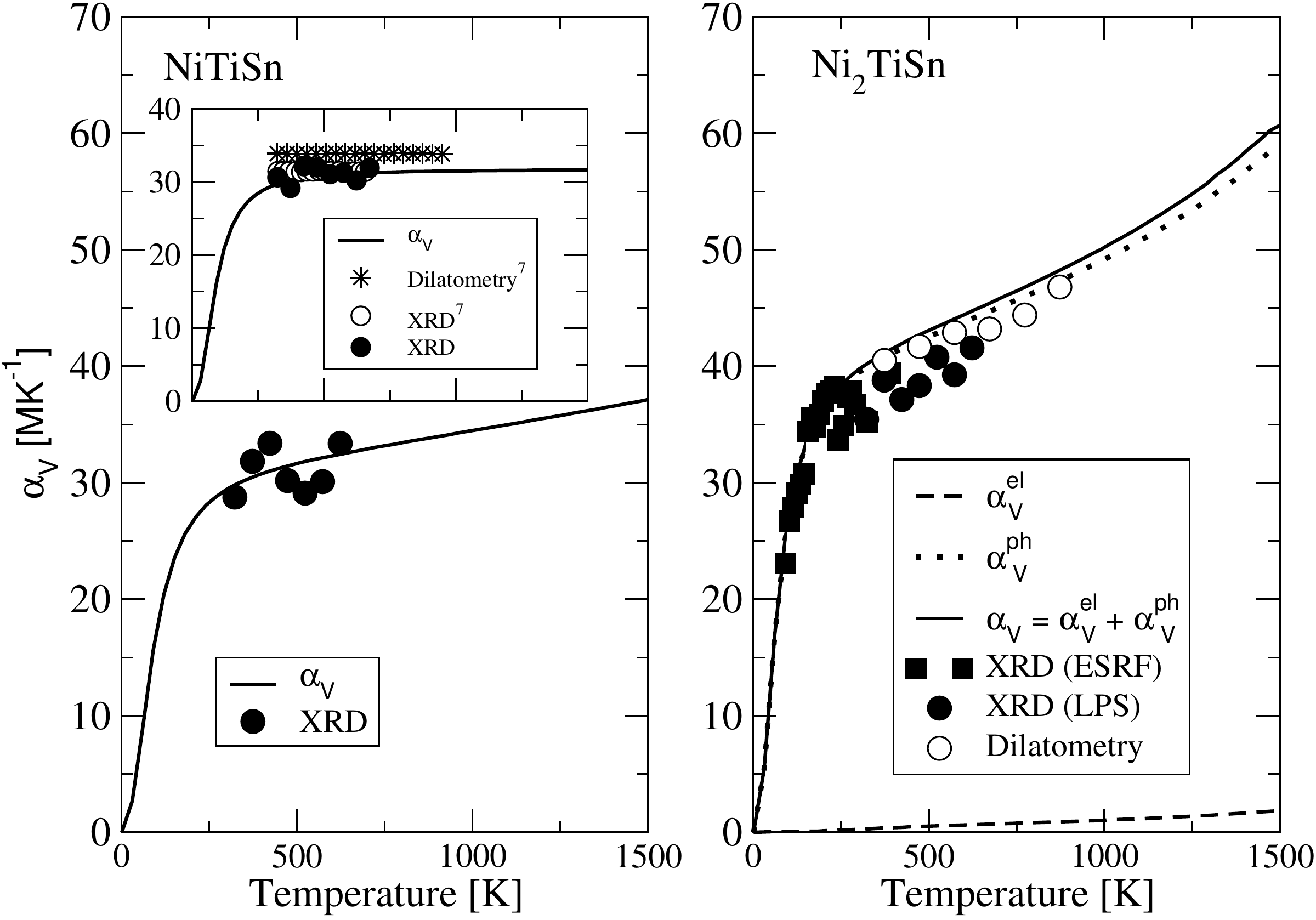}
\caption{Temperature dependence of the volumetric thermal expansion of NiTiSn and Ni$_2$TiSn obtained using Eq.~(5): comparison between XRD, dilatometry and calculations. 
In the inset of the left panel we compare our XRD and calculated values of NiTiSn obtained using Eq.~(12) to the experimental data of Jung {\it et al.}~\cite{Jung} also reported in their paper using Eq.~(12). Note: 1MK$^{-1}$ is equivalent to 1$\times$10$^{-6}$K$^{-1}$.}
\label{alpha}
\end{center}
\end{figure}

In \ref{alpha}, we show the temperature dependence of the volumetric
thermal expansion of NiTiSn and Ni$_2$TiSn. This thermal
expansion has been calculated using Eq. (7) with temperature-dependent
values of the bulk modulus and the unit cell volume obtained after
minimization of the free energy from a Birch-Murnaghan equation of state. The experimental thermal expansion was obtained from Eq. (5) with a derivative evaluated numerically using three consecutive points along the curve giving the experimental lattice parameter as a function of temperature. This methodology is expected to be more accurate (but less smooth) than a procedure involving a polynomial fit of this curve and then making the derivation, because it is independent of the degree of the polynomial used for the fit.\\
The decomposition of the thermal expansion into a vibrational and an electronic contribution is also reported on this figure for Ni$_2$TiSn. We observe that the vibrational contribution mainly dominates the thermal expansion while the inclusion of the electronic contribution leads to a minor correction up to 1500~K. In the case of NiTiSn (semiconducting compound), electronic contributions are negligible at the first order since the considered temperatures are significantly lower than the calculated energy band gap~\cite{patNiTiSn} ($E_g = $~0.49~eV).

NiTiSn has a lower thermal expansion coefficient than Ni$_2$TiSn in
the whole range of temperature due to its smaller Gr\"uneisen
parameter. This is also a consequence of the low phonon density
observed at low frequencies (below 100~cm$^{-1}$) in its vibrational
DOS with respect to Ni$_2$TiSn~\cite{patNiTiSn}. It is important to
bear in mind that the QHA used in the calculations of $\alpha_V$
neglects anharmonic effects, arising from phonon-phonon interactions,
which means that it can break down at high temperatures where such
effects can become significant. 
For Ni$_2$TiSn, the agreement is very good between the calculations
and the experiments in the recorded temperature range (80--873~K) even if the high temperature XRD measurements are slightly overestimated. The good agreement between experiments and simulations suggests that the QHA remains valid above room
temperature and at least up to 1000~K in this compound. The agreement is less impressive between our calculated and high temperature experimental values for NiTiSn even if the calculated curve goes through the experimental points. A possible 
explanation is that the phonon-phonon interactions (normal and umklapp processes)
are more important in this compound than in Ni$_2$TiSn. However, these interactions alone can not 
explain this lesser agreement. 
Indeed the NiTiSn sample is probably not monophasic because it is very
difficult experimentally to make high-quality samples, while the
synthesis of Ni$_2$TiSn is relatively easy and straightforward. Thus,
phonon-defect interactions should probably also be taken into account for
NiTiSn.

Concerning particular numbers, at room temperature our calculations
and dilatometry experiments give: $\alpha_V \simeq$ +40~MK$^{-1}$ for
Ni$_2$TiSn. For NiTiSn,  we find $\alpha_V \simeq$ +30~MK$^{-1}$
in the calculations while the experimental value (XRD) is slightly higher:
$\alpha_V^{exp} \simeq$ +31~MK$^{-1}$. At 750~K, the temperature at which the thermoelectric efficiency of NiTiSn is maximum \cite{ZT}, we obtain a value of $\alpha_V \simeq$ +33~MK$^{-1}$ close to the thermal expansion coefficient of iron at room temperature. The main errors in these calculated values of the thermal expansion are mainly related to: (i) the calculated equilibrium lattice constant that is slightly different from the experimental measurements (see \ref{BM}), (ii) the type of exchange--correlation functionals used in DFT, and (iii) the conditions where the QHA remains valid above room temperature.

\begin{figure}
\begin{center}
\includegraphics[width=15cm]{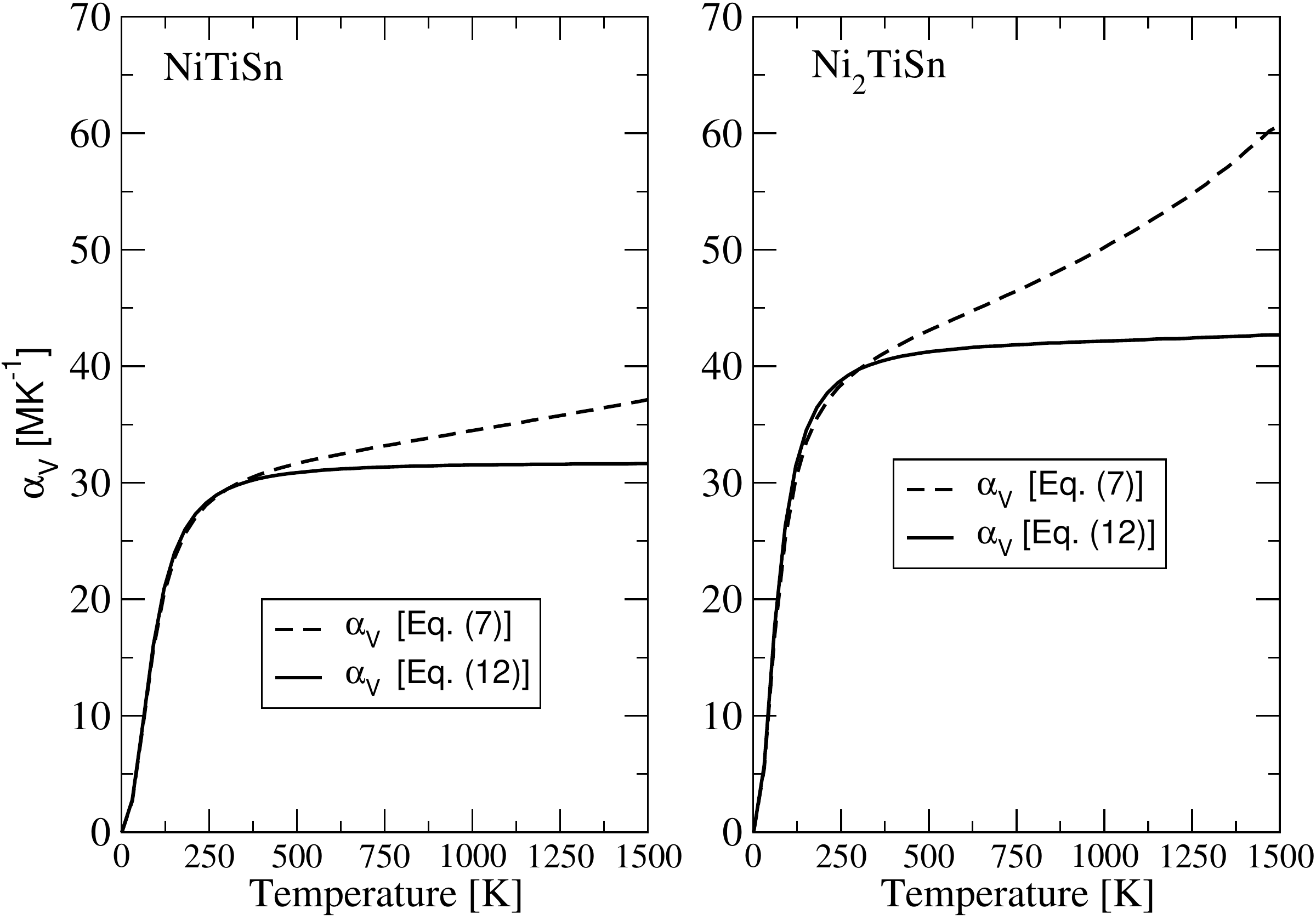}
\caption{Temperature dependence of the volumetric thermal expansion of NiTiSn and Ni$_2$TiSn calculated following Eq.~(7) (theoretical determination) and Eq.~(12) (experimental determination).}
\label{VV0}
\end{center}
\end{figure}

Note that in the literature, the average linear thermal expansion can also be defined by~\cite{Am}:
\begin{equation}
\langle\alpha\rangle_{ave}=\frac{1}{L_{RT}} \frac{L(T)-L_{RT}}{T-T_{RT}},
\end{equation}
where $L_{RT}$ is the sample length at the room temperature reference
and this definition was used to determine experimentally $\alpha_V$ of
NiTiSn by Jung {\it et al.}~\cite{Jung}. With this new definition, we
have calculated the thermal expansion of NiTiSn using Eq. (7) with
the values of the bulk modulus and unit cell volume fixed at 298~K. In
our calculations, these values are $B(298K)= 114$~GPa  and
$V(298K)=54.6$~\AA$^3$. 
In the inset of \ref{alpha} we compare the so-obtained calculated values with the
experimental results of Jung {\it et al.}~\cite{Jung} and with our own
experimental points determined with this equation. We observe an
overall good agreement between the three sets of experimental data and
the calculated curve. Using the same methodology for Ni$_2$TiSn for the calculated 
and experimental points, we found that
the thermal expansion of both compounds grows rapidly up to 350~K and
then becomes nearly constant above this temperature. Thus, the main
differences between the thermal expansion calculated using Eq. (7) and
Eq. (12) appear roughly above room temperature (see \ref{VV0}).
\subsection{D. Heat capacities} 
\begin{figure}
\begin{center}
\includegraphics[width=15cm]{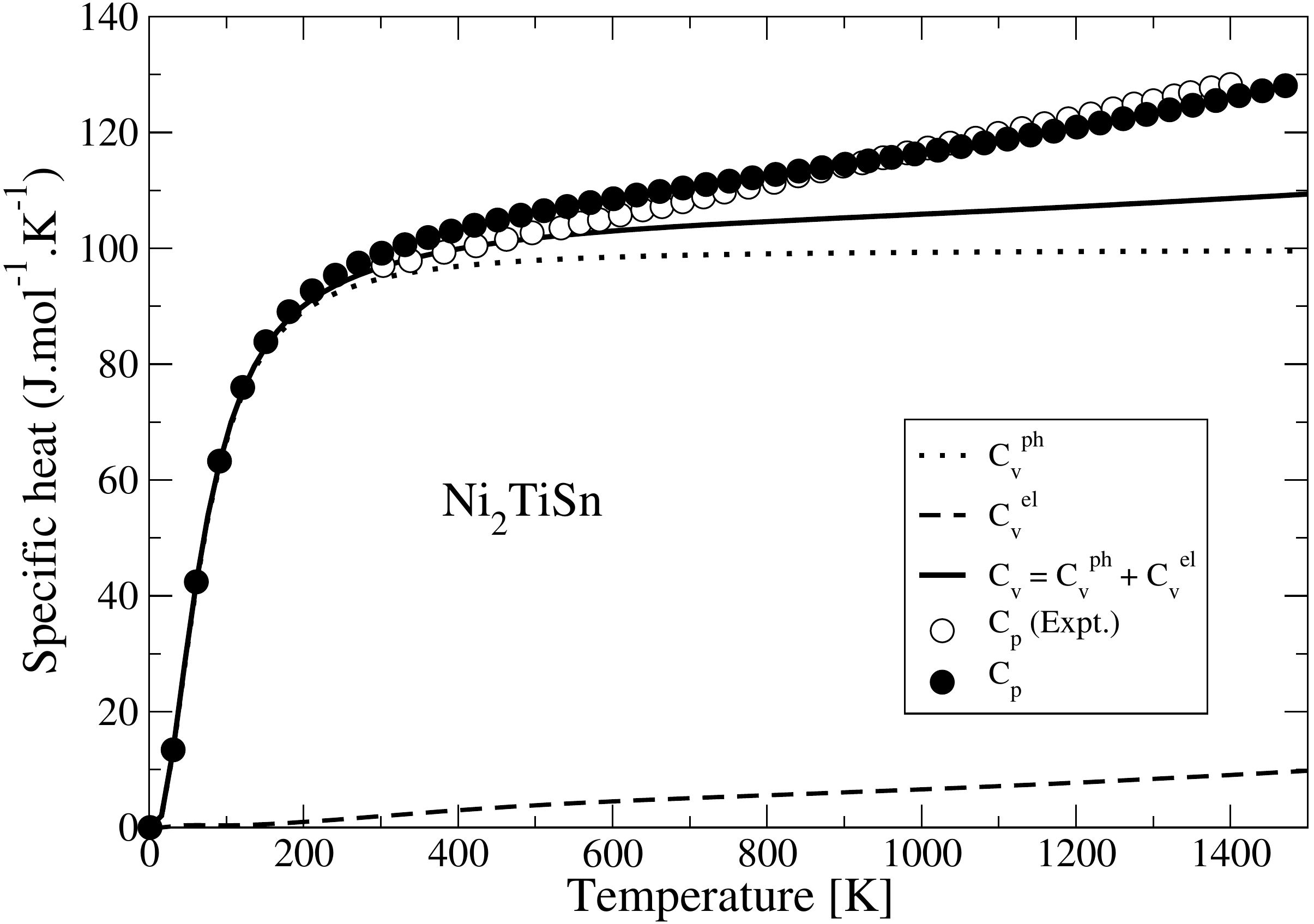}
\caption{Temperature dependence of the Ni$_2$TiSn specific heat at constant-volume ($C_v$) and constant-pressure ($C_p$). Electronic ($C_v^{el}$) and phonon ($C_v^{ph}$) contributions to the specific heat at constant-volume are also reported. The open circles are the experimental values from Nash~\cite{PPT}.}
\label{cvphel}
\end{center}
\end{figure}

\begin{figure}
\begin{center}
\includegraphics[width=15cm]{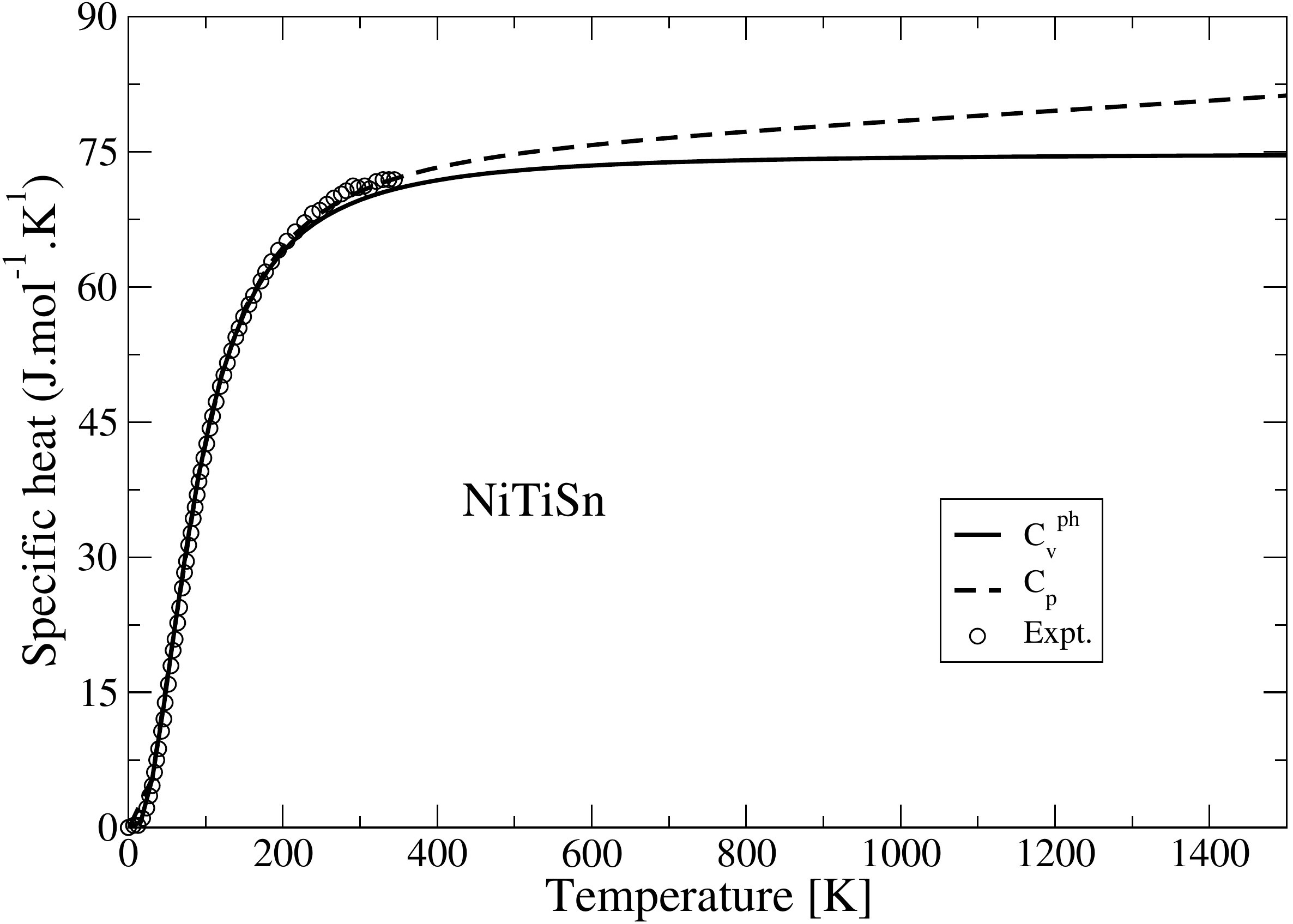}
\caption{Constant volume ($C_v^{ph}$, vibrational part), and constant pressure ($C_p$) specific heat of NiTiSn versus temperature. The open circles are the experimental values from Zhong ~\cite{Zhong}.}
\label{cvcp}
\end{center}
\end{figure}

Once the phonon spectrum over the whole Brillouin zone is available, the vibrational heat capacity at constant-volume ($C^{ph}_v$) can be calculated by \ref{cvph}, while the electronic contribution to the heat capacity at constant-volume ($C^{el}_v$) can be obtained from the electronic DOS by using \ref{cvel}. The temperature dependence of the specific heats at constant-pressure $C_p$ of both compounds have been calculated using \ref{cpcv}. The different heat capacities and contributions are plotted in \ref{cvphel} in the case of Ni$_2$TiSn, whereas only the vibrational contribution to $C_v$ is considered for NiTiSn in \ref{cvcp}. First, the $C^{ph}_v$ contributions tend to the classical Dulong and Petit constant as the temperature increases, $C^{ph}_v=3Nk_B$ where $N$ is the number of atoms, while $C^{el}_v$ and $C_p$ still increase. In particular, $C_v^{ph}(T\rightarrow\infty)= $99.77~J.mol$^{-1}$.K$^{-1}$ for Ni$_2$TiSn and $C_v^{ph}(T\rightarrow\infty)= $74.83~J.mol$^{-1}$.K$^{-1}$ for NiTiSn. Then, for the thermal electronic contributions to the specific heat of Ni$_2$TiSn, we find that $C^{el}_v$ is not negligible at high temperatures though significantly smaller than $C^{ph}_v$. This character can be understood from the high electronic DOS observed near the Fermi level~\cite{colinet}. For NiTiSn, the small difference between $C_p$ and $C_v$ at high temperatures is a consequence of its smaller thermal expansion. In contrast to the high temperature range, the difference between $C_p$ and $C_v$ can be neglected for both compounds at low temperature (below $\approx$ 300~K). For Ni$_2$TiSn, we observe for $C_p$ a remarkable agreement between our calculated values and the experimental~\cite{PPT} ones above room temperature. This highlights the reliability of the calculation of the thermal expansion and the validity of the QHA above room temperature in this compound. Below this temperature, only measurements of the specific heat below 25~K have been reported in the literature to our knowledge~\cite{Aliev,Boff}, leaving the 25--300~K range unexplored. In the case of NiTiSn, this agreement is also satisfactory with the experimental data obtained by Zhong~\cite{Zhong} up to 400~K. There is no experimental values reported in the literature above room temperature.

\section{V. CONCLUSIONS}
We have determined {\it via} first principles calculations and experimental measurements including three different methods (high energy X-Rays (ESRF), standard X-Rays (Montpellier) and dilatometry) the thermal characteristics of NiTiSn and Ni$_2$TiSn. First, the calculated dispersion curves of the mode Gruneisen parameter are positive throughout the Brillouin zone, suggesting that these compounds undergo a regular thermal expansion. No clear dominant mode Gruneisen parameter is observed for NiTiSn whereas some branches have a high Gruneisen parameter at the zone center for Ni$_2$TiSn. As a consequence, the overall Gruneisen parameter of Ni$_2$TiSn is the highest. Then, the calculated thermal expansion coefficients are in excellent agreement with the experimental values in the recorded temperature range (80-873~K) for Ni$_2$TiSn. In the case of NiTiSn, this agreement is less impressive at high temperatures. These deviations can be attributed to stronger phonon-phonon interactions that are neglected within the QHA, but also to phonon-defect interactions due to the low quality of the sample. The constant-volume heat capacity has been calculated for both compounds and compared with the experimental data reported in the literature. In particular, we decomposed that of Ni$_2$TiSn into a purely electronic  and a phonon-mediated contribution, and we discussed each of them. Finally, we derived their constant-pressure heat capacities. The remarkable agreement found for Ni$_2$TiSn with the experimental data highlights the reliability of the calculation of its thermal expansion and the validity of the QHA above room temperature in this compound. This exhaustive study of the thermal properties of NiTiSn and Ni$_2$TiSn is the first step in the determination via {\em ab initio} methods of the thermal conductivity of these Heusler materials. Such completely first principles methods do not exist yet even though obtaining reliable thermal conductivities is fundamental to predict new and efficient thermoelectric materials.


\section*{Acknowledgements}
We are grateful to Bernard Fraisse, Dominique Granier (ICGM
Montpellier) for NiTiSn X-ray measurements , Pierre Antoine Albouy
(LPS, Orsay) for Ni$_2$TiSn X-ray measurements and Laurent Duffours
(PRIME Verre) for fruitful discussions on the dilatometry
measurements. We thank the ICGM for financial support. 
Part of the simulations have been performed at the National computer
center CINES in Montpellier. PGY and FS acknowledge the European
Synchrotron Radiation Facility for allowing beam time and Dr. D. Chernyshov for fruitful discussions.



\clearpage

\clearpage

\clearpage

\clearpage

\clearpage

\clearpage


\begin{thebibliography}{00}
\bibitem{Note} In cubic structure, the volumetric ($\alpha_V$) and the linear ($\alpha_L$) thermal expansion coefficients are proportional according to: $\alpha_V=3\alpha_L$.
\end{thebibliography}
\end{document}